\documentclass[acmsmall,screen,nonacm]{acmart}
\acmSubmissionID{70}

\usepackage{url}
\usepackage{listings}
\usepackage{caption}
\usepackage{subcaption}
\usepackage{cleveref}
\usepackage{graphicx}
\usepackage{wrapfig}

\lstdefinestyle{mystyle}{
    basicstyle=\ttfamily\footnotesize,
    breakatwhitespace=false,
    breaklines=true,
    captionpos=b,
    keepspaces=true,
    numbers=left,
    numbersep=5pt,
    showspaces=false,
    showstringspaces=false,
    showtabs=false,
    tabsize=2,
    xleftmargin=0.4cm,
    commentstyle=\color{olive},
}

\begin{document}
\title{Partial Evaluation, Whole-Program Compilation}
\subtitle{\textit{Or,} We Have a Compiler at Home}
\author{Chris Fallin}
\email{chris@cfallin.org}
\orcid{0000-0002-6733-1803}
\affiliation{%
  \institution{Fastly}
  \city{San Francisco}
  \state{California}
  \country{USA}
}
\author{Maxwell Bernstein}
\email{acm@bernsteinbear.com}
\orcid{0000-0003-3130-7059}
\affiliation{%
  \institution{Recurse Center}
  \city{Boston}
  \state{Massachusetts}
  \country{USA}
}

\newcommand\todo[1]{\textcolor{red}{[TODO: #1]}}
\newcommand\ignore[1]{}

\newcommand\weval{weval}
\newcommand\wevaled{wevaled}
\newcommand\wevals{weval's}
\newcommand\wevaling{wevaling}

\begin{abstract}
There is a tension in dynamic language runtime design between speed
and correctness: state-of-the-art JIT compilation, the result of enormous
industrial investment and significant research, achieves heroic
speedups at the cost of complexity that can result in serious correctness
bugs. Much of this complexity comes from the existence of multiple
tiers and the need to maintain correspondence between these separate
definitions of the language's semantics; also, from the indirect
nature of the semantics implicitly encoded in a compiler backend.  One
way to address this complexity is to automatically derive, as much as
possible, the compiled code from a single source-of-truth; for
example, the interpreter tier. In this work, we introduce a partial
evaluator that can derive compiled code ``for free'' by specializing
an interpreter with its bytecode. This transform operates on the
interpreter body at a basic-block IR level and is applicable to almost
unmodified existing interpreters in systems languages such as C or
C++. We show the effectiveness of this new tool by applying it to the
interpreter tier of an existing industrial JavaScript engine,
SpiderMonkey, yielding $2.17\times$ speedups, and the PUC-Rio Lua
interpreter, yielding $1.84\times$ speedups with only three hours'
effort. Finally, we outline an approach to carry this work further,
deriving more of the capabilities of a JIT backend from first
principles while retaining semantics-preserving correctness.

\end{abstract}

\maketitle

\section{Introduction}
\label{sec:intro}

Most dynamic language runtimes start as interpreters, for their
numerous initial advantages: interpreters are easier to develop and
extend than compilation-based alternatives; they are likewise easier
to debug; and they are usually more portable, relying less on
platform- or ISA-specific details to generate and execute code. Over
time, dynamic language runtimes tend to build run-time type profiling
and code specialization features, and going further, some develop
just-in-time (JIT) compiler backends to remove interpretation
overhead. A JIT is effective but not free: it is a second
implementation of language semantics that may diverge in hard-to-debug
ways from the corresponding interpreter, it generates specialized code
that may depend on invariants that can be invalidated at run-time, and
this generated code is ephemeral and thus harder to
audit.

Security-conscious platforms often eschew run-time code generation for
this reason. For example, iOS does not permit it outside the built-in
web engine (and turns it off in this engine too in lockdown mode), and
the Edge browser has a secure mode that disables
JIT~\cite{edge-superdupersecure}. The three major JavaScript engines
(V8, SpiderMonkey and JSC) regularly have CVEs resulting from subtle
JIT bugs (e.g.~\cite{cve-2024-4761} as one recent example in V8). Even
coarse-grained sandboxing, such as V8's
``ubercage''~\cite{v8-ubercage}, does not protect against correctness
bugs that can violate isolation when multiple server-side tenants
share one VM via isolates~\cite{cloudflare-v8-isolates}, or when
multiple requests with different users' data are processed by one
engine. Separately, the continued prevalence of JS engines' security
bugs indicates that full language-semantics correctness is difficult
to maintain; and, even if a bug does not yield a sandbox escape, it
can be catastrophic to individual applications when miscompilations
alter the application's logic.

We thus see a tension between the ever-increasing need for efficient
execution of dynamic languages -- manifested in the enormous
engineering investments in JIT compilers and language runtime
optimization -- and the need for security and correctness as these
languages are used to implement the underpinnings of modern
infrastructure.

In this paper, we explore a technique -- partial evaluation -- to
derive a language runtime's compiler backend automatically from an
interpreter that already exists. This is an old technique, going back
to Futamura~\cite{futamura71,futamura99}, with numerous modern
implementations~\cite{graal-pldi2017,pypy,deegen,sempy,lua-aot}.  We
observe, however, that most extant tools today require the interpreter
to be \emph{developed for the purpose} -- in other words, written in
terms of a framework (as with Graal/Truffle~\cite{graal-pldi2017}), or
in terms of a semantic-definition DSL~\cite{deegen,sempy}, or in a
special restricted language~\cite{pypy}. Our contribution is to design
a partial-evaluation transform that works on a mostly unmodified
interpreter body, at the IR level: it transforms an arbitrary
control-flow graph of basic blocks, in SSA form, unrolling an
interpreter loop as a side-effect that falls out of a general
``context specialization'' mechanism (\S\ref{sec:weval}). Our tool is
an open-source, industrial-strength compiler that we call \weval{}. In
its initial form, it processes interpreters compiled to
WebAssembly~\cite{wasm-pldi17}, as a portable and easily transformable
IR that many compiler toolchains can target, but without loss of
generality it should work on any optimizing compiler IR that uses basic
blocks, such as LLVM~\cite{llvm}, with some fairly minimal requirements
(\S\ref{sec:weval-ir-requirements}).

This approach provides several benefits. First, it allows easy, rapid
provisioning of a compiler-based backend for a language runtime, as we
show in our case study on Lua (\S\ref{sec:casestudy-lua}) where we
managed to achieve a speedup of $1.84\times$ with three hours'
effort. There exists a long tail of language implementations that have
only interpreters and that could benefit from this technique. Even
established language runtimes can benefit on new platforms not
supported by existing their JIT backends: for example, the
SpiderMonkey JS engine (\S\ref{sec:casestudy-spidermonkey}) has no JIT
tiers when compiled to run on a server-side Wasm platform
(i.e. sandboxed within a Wasm module), whereas \weval{} allows us to
attain a $2.17\times$ speedup on average ``for free'' -- deriving the
result from exactly the same interpreter source. Second, it provides a
realistic pathway toward single-source-of-truth definitions of language
semantics. We describe a plausible future path for carrying this
approach forward to include profile feedback in a
\emph{semantics-preserving} way, showing how one might derive a
competitive JIT from language semantics (in an interpreter) from first
principles.

\section{Futamura Projections and Partial Evaluation}
\label{sec:background-futamura}

In this work, we observe that we can \emph{automatically} produce
compiled code from an interpreter body and its interpreted program
input. In order to understand this further, we first need to
understand how to automatically produce compiled code from an
interpreter: that is the \emph{Futamura projection}.

\subsection{The Futamura Projection}

Futamura~\cite{futamura71,futamura99} introduced the concept of
partial evaluation in the context of compilation: by \emph{partially
evaluating} an interpreter with its interpreted program, we obtain a
compiled program. Consider an interpreted program execution as a
function invocation, where the interpreter receives two inputs, the
interpreted program and the input to that program:
$\textit{Interp}(\textit{Prog}, \textit{Input})$. The key idea of the
first Futamura projection is to substitute in a constant $C$ for the
$\textit{Prog}$ argument, yielding a new function that we can consider
a compiled form of the user program. Then we have
$\textit{Compiled}(\textit{Input}) = \textit{Subst}_{ \textit{Prog}=C}
(\textit{Interp}(\textit{Prog}, \textit{Input}))$.  What we have
described so far is the \emph{first} Futamura projection: it is the
partial evaluation of the interpreter with an interpreted program,
yielding a compiled program. Futamura also defines the second and
third projections: the second projection enhances compilation speed,
and the third produces a compiler-compiler tool, but both are less
tractable than the first projection, and we will not describe them
further.

\subsection{Optimizing Compilation: Interpreted Program to Specialized Code}

One could achieve a basic kind of compilation by joining an
interpreter with a snapshot of its input (bytecode), perhaps by
linking the interpreter with an additional data section and some
startup code. This fits the definition of a Futamura projection in a
trivial sense. However, practically speaking, this ``compilation''
lacks many of the properties one usually expects from compiled code.
Mainly this relates to performance: the combined module retains the
performance characteristics of the interpreter, because the
interpreter body is unchanged. Let us now state a definition that sets
a minimum bar for a compilation with the desired performance:

\textbf{Definition.} A partial evaluator performs a
\textbf{bytecode-erased compilation} if the resulting specialized
program does not load data from the bytecode stream; hence it
should have program points that statically correspond to the
\emph{interpreted} program rather than the \emph{interpreter}, and
should no longer dynamically dispatch behavior based on the original
bytecode.\footnote{This concept is very similar to
Jones-optimality~\cite{jones-peval-1990}, which specifies that a
partial evaluator should ``remove all computational overhead caused by
interpretation.''}

\begin{wrapfigure}[18]{r}{0.5\textwidth}
\begin{lstlisting}[language=C]
void interpret(bytecode_t* pc,
               Value* stack) {
  while (true) {
    switch (*pc++) {
    case OP_add:
      Value v1 = *stack++;
      Value v2 = *stack++;
      *--stack = value_add(v1, v2);
      break;
    /* ... other opcodes ... */
    }
  }
}
\end{lstlisting}
\caption{An sketch of an interpreter loop written in C.}
  \label{fig:motivation-interp-loop}
\end{wrapfigure}

That is, we consider the result a desirable compilation if it replaces
the interpreter's control flow with the native control-flow graph of
the interpreted program. This is both itself a speedup (in our
observations, often 1.5-2x) and a substrate for further optimizations:
each instance of an opcode becomes its own static code, we can
optimize it separately, and together with the opcodes around it.

\subsection{Optimizing an Interpreter with its Input}
\label{sec:motivation}

The key question is: how can we practically \emph{expand bytecode to
specialized code} by partially evaluating an interpreter loop? As we
\begin{wrapfigure}[14]{r}{0.45\textwidth}
\begin{lstlisting}[language=C]
void interpret_specialized_func0(
    bytecode_t* _pc_unused,
    Value* stack) {
  Value v1 = stack[0];
  Value v2 = stack[1];
  stack[1] = value_add(v1, v2);
  return;
}
\end{lstlisting}
\caption{Compiled code resulting from constant propagation of
  \texttt{interpret} from Fig.~\ref{fig:motivation-interp-loop} on one
  opcode.}
\label{fig:motivation-compiled-one-opcode}
\end{wrapfigure}
will see in the rest of this section, there are various design points,
requiring various compromises in the way that the interpreter is
expressed.

Above we introduced an algebraic analogy to partial evaluation,
namely, substituting a variable for a constant value and simplifying
(optimizing). What happens if we apply the analogous compiler analysis
and transform, namely constant propagation and folding?

Consider the body of the interpreter loop in
Fig.~\ref{fig:motivation-interp-loop}. If we take a function
\texttt{func0} of a single opcode, say \texttt{OP\_add}, and we take a
constant initial stack pointer offset, we might imagine taking the
body of the interpreter and producing code similar to
Fig.~\ref{fig:motivation-compiled-one-opcode}.

This code results because constant propagation can convert the fetch
of the opcode to its constant value \texttt{OP\_add}. This in turn
works because we are processing a partial evaluator invocation in
which the user has promised that this memory is constant (``specialize
this function when this pointer points to this data''). The
specializer can then branch-fold the \texttt{switch} to the one case
actually taken due to the constant selector, and constant-fold the
offsets from \texttt{stack}.

\begin{figure}
  \centering
  \includegraphics[width=0.8\textwidth]{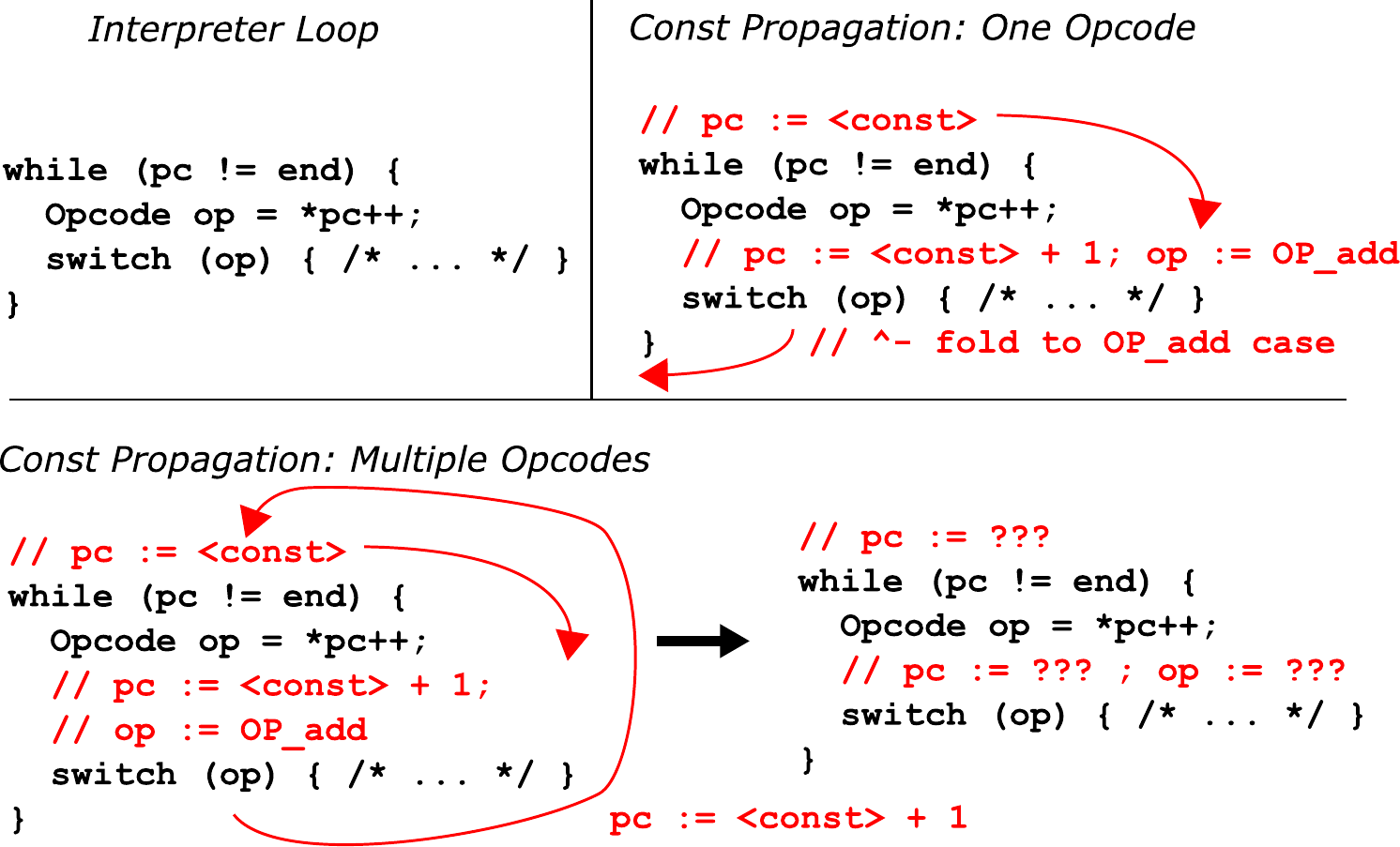}
  \caption{An illustration of constant propagation over an interpreter
    loop: with one iteration, we can deduce constant values, but
    multiple iterations cause the analysis to degrade to ``unknown''
    because all iterations are considered together.}
  \label{fig:motivation-moap}
\end{figure}

However, as soon as we advance to a program of \emph{two} opcodes --
before even considering control flow within the interpreted program --
we run into issues with constant propagation. In fact, we glossed over
the issue in the single-opcode example: how do we handle the
interpreter loop backedge? A classical iterative dataflow analysis,
such as constant propagation, computes a Meet-Over-All-Paths
solution~\cite{dragonbook}, meaning that it produces one analysis
conclusion per \emph{static program point}, merging together all paths
that could reach that point. At the top of the interpreter loop (one
program point), what is \texttt{pc}? When we merge all iterations
together, we only reach the conclusion that it is not constant,
because we are analyzing all opcodes at once. The rest of the
interpreter then fails to specialize to the bytecode: \texttt{pc} is
not constant, so neither is \texttt{*pc}, so we cannot branch-fold the
\texttt{switch}, so the result of specialization is only a copy of the
original interpreter, with nothing changed. This situation is
illustrated in Fig.~\ref{fig:motivation-moap}.

The heart of the issue is that in order to compile the bytecode to
target code, we need to somehow iterate over the bytecode operators
and emit code for each one, and this iteration happens \emph{at
compile-time}. A constant-folding pass that retains the original CFG,
only substituting in constants where known, will not lead to this
output. Can we build an analysis that somehow knows how to
\emph{unroll arbitrary interpreter loops over the bytecode}?

One possible approach is to \emph{unroll all loops} by analyzing the
interpreter along a \emph{trace}: in other words, discarding the
Meet-Over-All-Paths principle. This approach is appealing in its
simplicity. However, it can result in unbounded work: thus, it must be
limited by trace size or some other metric, and it can have surprising
worst-case cost.

A more targeted approach, taken by PyPy~\cite{pypy}, is to detect
\emph{hot loops} that result from loops in the interpreted program,
and trace the interpreted execution of these loops, instantiating and
specializing the interpreter loop once for each opcode in the loop.

This approach resolves the above limits but has as a
\emph{profile-driven} approach, it requires execution of the
interpreted program before compilation can commence. In some settings,
we may desire fully ahead-of-time compilation, or we may not have
adequate or representative test inputs for the interpreted program (or
may not be able to run it at all in the compiler's execution
environment, if it has other dependencies). Additionally, it can
suffer from brittle \emph{performance cliffs}: if the control-flow
path during run-time diverges from that seen during compilation,
execution must revert to the interpreter (possibly ameliorated by
attaching ``side traces'' over time). This behavior of ``falling off
the trace'' was a well-known failure mode in the TraceMonkey
JavaScript JIT~\cite{tracemonkey}.

The downside of both of the above options is that, in attempting to
specialize an interpreter loop fully automatically, they rely on
heuristics that can fail fairly easily. One alternative is to allow --
or indeed, require -- the interpreter author to \emph{explicitly
denote the interpreter loop} and how it should be specialized.
The high-level idea would be to devolve control of the main
interpretation loop to a \emph{framework} that the partial evaluator
is somehow aware of. This framework would understand the format of the
interpreted program (e.g., bytecode or AST nodes), and would take care
of dispatching to implementations of opcode semantics provided by the
interpreter author. In this way, the partial evaluator could directly
translate the bytecode or other interpreted program representation to
compiled form by copying over and concatenating the implementations of
each opcode.

While this ought to work robustly, because the partial evaluator is
co-designed and developed with the interpreter framework, it has the
major disadvantage that it requires the interpreter to be written in a
way specific to this tool. An existing interpreter is likely to be
difficult to port to this framework.

\section{The \weval{} Transform: User-Context-Controlled Constant Propagation}
\label{sec:weval}

We have argued that to produce a \emph{bytecode-erased compilation} --
that is, to produce compiled code that has a separate program point
for each bytecode, turning the transitions in interpreter state into
true control-flow edges -- we need to somehow \emph{unroll the
interpreter loop} during partial evaluation, analyzing the loop body
separately for each interpreted-program operator. Furthermore, we wish
to do this without rewriting the interpreter to conform to a framework
that understands the structure of the interpreted program. Rather, we
want to support an existing interpreter, with minimal modifications,
using its own logic to ``parse'' the bytecode as we translate it
opcode by opcode. In this section, we will introduce a transform that
does exactly this. We call this the \weval{} transform, short for
``WebAssembly [partial] evaluator.''

The transform operates on a function body represented as a
control-flow graph (CFG) of basic blocks in static single assignment
(SSA) form. Due to the problem-space that we built this tool to
address (see \S\ref{sec:casestudy-spidermonkey}), we build and use a
framework that allows for SSA CFG-based Wasm-to-Wasm
compilation. However, without loss of generality, this transform can
apply to any IR that is a CFG of basic blocks, such as
LLVM~\cite{llvm} (\S\ref{sec:weval-ir-requirements}). This transform
is relatively small for its power, measuring at 5KLoC of Rust.

\subsection{Key Idea \#1: User Context}
\label{sec:weval-context-intrinsics}

Recall that we began our discussion of the Futamura projection by
noting how constant propagation addresses the problem fully in the
single-opcode case, but fails as soon as more than one opcode exists
in the interpreted program
(Fig.~\ref{fig:motivation-moap}). Specifically, when the
constant-propagation analysis follows the interpreter backedge, the
``next'' value of the interpreter program counter conflicts with the
previous value, and we conclude that nothing is constant at all.

To address this, we allow the interpreter to \emph{selectively
introduce context specialization via intrinsics} to separate the
analysis of each interpreter loop iteration. The intrinsic invocation
appears like \texttt{update\_context(pc)} at some point before the
loop backedge, as shown in Fig.~\ref{fig:weval-annotations}; when
performing an iterated dataflow analysis for constant propagation,
this causes analysis to flow to successor blocks in a \emph{new
context}. In other words, the set of program-point locations analyzed
by iterated dataflow analysis is dynamic and expandable. This will be
illustrated by an example below in
Fig.~\ref{fig:weval-worked-example}.

This annotation is lightweight and minimal, yet it unlocks an entire
specialization pipeline: it drives code duplication exactly and only
where needed to replicate the interpreter body according to the
overall schema of the interpreted bytecode, and the rest of the
specialization falls out of this. In other words, by avoiding the
``meet-over-all-paths'' trap that we described in
\S\ref{sec:motivation}, we achieve a \emph{bytecode-erasing
compilation} that produces an output control-flow graph that follows
the bytecode rather than the interpreter.

\begin{wrapfigure}[18]{l}{0.45\textwidth}
  \begin{lstlisting}[language=C]
 void interpret(bytecode_t* pc) {
  while (true) {
    switch (*pc++) {
      /* ... */
    }
    // Update analysis context:
    // backedge reaches loop
    // header in a new context,
    // maintaining constantness
    // of pc.
    update_context(pc);
  }
}
\end{lstlisting}
  \caption{Annotations to context-specialize analysis of an
    interpreter function.}
  \label{fig:weval-annotations}
\end{wrapfigure}

Note that context may be nested: our actual intrinsics include
\texttt{push\_context()} and \texttt{pop\_context()}, allowing, e.g.,
value-specialization or manual loop unrolling to occur inside of the
main interpreter loop unrolling.

Furthermore, note that this requires the context value (\texttt{pc}
here) to be a known \emph{constant} at specialization time. This will
be the case for bytecode-driven control flow with a fixed CFG, but,
e.g., an opcode that computes an arbitrary bytecode destination would
not be workable. (What CFG should result in the compiled code? Will
there be an edge to every block?) To support computed-gotos with a
known list of destinations (e.g., switches or exception handlers), we
allow for value specialization
(\S\ref{sec:weval-value-specialization}).

Finally, note that this intrinsic is not load-bearing for correctness: it
splits constant-propagation context, but the \weval{} transform is sound
regardless of how many separate contexts are used to analyze duplicates of
code. The worst that happens with an arbitrarily wrong context is that
specialization collapses back to ``nothing is known'' and the result is the
original interpreter body. Separately, we provide an intrinsic that
\emph{asserts compile-time constantness} to help debug such \emph{performance}
issues.

\subsection{Key Idea \#2: Context-Specialized Code Duplication}
\label{sec:weval-transform}

Given a \emph{generic} function to be \emph{specialized} with a set of
constant parameters, we now define the worklist-driven algorithm that
produces the specialized function body.

The algorithm operates over the generic (input) function in an
SSA-based IR containing basic blocks, and is driven by a worklist of
blocks to specialize. Blocks in the generic function are specialized
\emph{per context} into blocks in the specialized function. We keep a
mapping from $\langle$basic block, context$\rangle$ tuples to
specialized blocks, and likewise for SSA value numbers. We specialize
one block at a time, performing constant-propagation analysis,
const-folding and branch-folding as we flow forward. We track and
update the \emph{current context} as flow-sensitive analysis state,
updated as necessary by intrinsics. When we reach a branch
instruction, look up target block(s) in the current context. If
already processed, create an edge to the corresponding specialized
block. Otherwise, enqueue the block in context on a worklist. We
provide this algorithm as pseudocode in Fig.~\ref{fig:weval-algo}.

\begin{figure}
  \centering
\begin{lstlisting}[language=Python,basicstyle=\tiny\ttfamily]
worklist = []    # Worklist of (Ctx, Block)
blockmap = {}    # Map from (Ctx, Block) to SpecializedBlock
valuemap = {}    # Map from (Ctx, Value) to SpecializedValue
valuestate = {}  # Map from SpecializedValue to CpropAnalysisState
# ... Dependency management to re-enqueue blocks omitted
# ... Flow-sensitive state management omitted

def partially_evaluate(func, args):
  for (arg, cprop_state) in args:
    valuestate[arg] = cprop_state

  initial_ctx = create_root_context()
  worklist.append((initial_ctx, func.entry_block))
  while len(worklist) > 0:
    (ctx, block) = worklist.pop()
    partially_evaluate_block(func, ctx, block)

def partially_evaluate_block(func, ctx, block):
  specialized_block = blockmap[(ctx, block)] or create_new_block(ctx, block)
  # We may be revisiting due to updated abstract state; empty the block.
  clear_block(specialized_block)

  # Partially evaluate and transcribe over the instructions.
  for inst in func[block]:
    specialized_args = [valuemap[(ctx, value)] for value in func.inst_args(inst)]
    (specialized_inst, abstract_state, ctx) = partially_evaluate_inst(
        func, ctx, specialized_block, inst, specialized_args,
        [valuestate[spec_arg] for spec_arg in specialized_args])
    valuestate[specialized_inst] = abstract_state
    append_to_block(specialized_block, specialized_inst)

  # Evaluate the terminator (branch) targets, enqueueing more blocks.
  # Omitted: if conditional or switch and constant selector, branch-fold.
  set_terminator(specialized_block,
    [evaluate_target(ctx, target) for target in terminator_targets(func, block)])

def partially_evaluate_inst(func, ctx, specialized_block,
                            inst, specialized_args, abs_states):
  if inst is intrinsic 'update_context':
    ctx = abs_states[0]  # Assert this is a constant; we do not support runtime ctx
  elif ...: # Handle other intrinsics
  else:
    abstract_state = constant_propagate(inst_opcode(func, inst), abs_states)
    if abstract_state is constant c:
      inst = create_const_value(specialized_block, c)
    else:
      inst = clone_inst(inst, specialized_block)
  return (inst, abstract_state, ctx)

def evaluate_target(ctx, target):
  specialized_target = blockmap[(ctx, target)] or create_new_block(ctx, target)
  # ... Meet flow-sensitive state into entry state (omitted) ...
  if newly created or entry state changed:
    worklist.push((ctx, target))
  return specialized_target
\end{lstlisting}
  \caption{Pseudocode for the main specialization (Futamura
    projection) algorithm.}
  \label{fig:weval-algo}
\end{figure}

\begin{figure}
  \centering
  \includegraphics[width=0.7\textwidth]{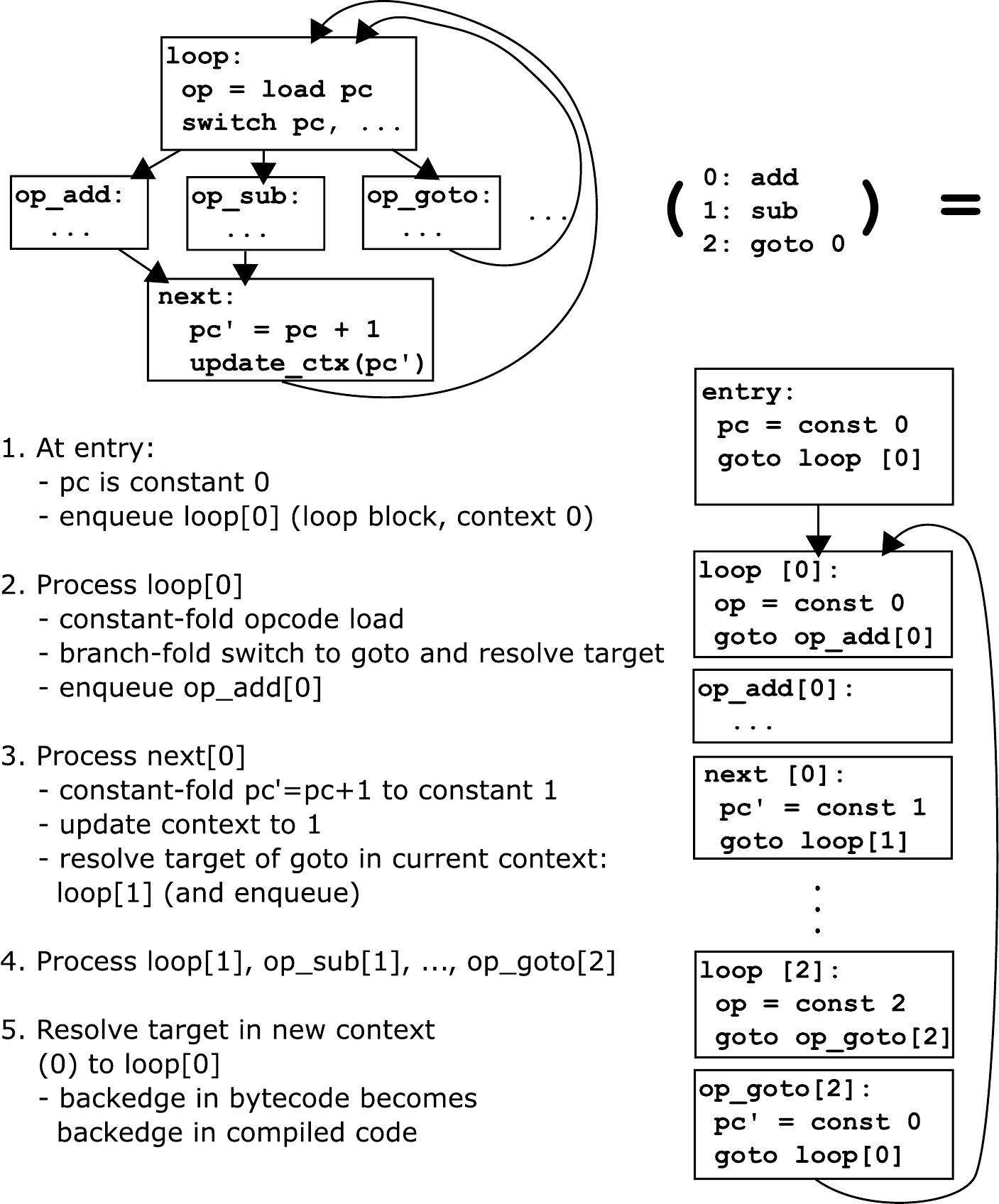}
  \caption{An example of a partial evaluation of a simple interpreter
    on a three-opcode interpreted program.}
  \label{fig:weval-worked-example}
\end{figure}

In Fig.~\ref{fig:weval-worked-example} we show an example of a
specialization of a simple interpreter (supporting three opcodes, ADD,
SUB and GOTO) for a bytecode program that performs ADD and SUB
operations in an infinite loop. The interpreter is annotated with
context updates, and the specialization provides the semantic
information that the bytecode is constant. Note, however, that no
other knowledge of \emph{interpreters}, per-se, is needed: this is a
fully general transform for duplicating and constant-specializing
code.

The analysis is worklist-driven and runs until fixpoint, but in
practice in most cases, makes one pass over the bytecode, emitting the
portion of the interpreter-switch corresponding to each opcode. That
is, the overall scheme of a single-pass template compiler \emph{falls
out automatically}, without us having to adapt the interpreter or
bytecode into a framework that understands this flow.

The resulting compiled code contains a control-flow graph that
corresponds to the \emph{interpreted program}, with its loop (the JMP
backedge), rather than the interpreter. Thus, we have a
\emph{bytecode-erased compilation} as a result of a partial
evaluation. This is an instance of a first Futamura projection.

\subsection{Key Idea \#3: Directed Value-Specialization}
\label{sec:weval-value-specialization}

Basic block specialization requires \emph{compile-time constant}
context values: otherwise, we cannot resolve branch targets to blocks
in the specialized function statically.

However, an interpreted program will naturally have
run-time-data-dependent control flow in the form of conditional
branches. An interpreter will implement these branch opcodes either
with its own branch, conditionally updating its ``next PC'' value, or
with a branchless conditional-select (e.g., \texttt{condition ?
  targetPC : fallthroughPC}). The issue with both of these is that
control flow reconverges to a single backedge to the next interpreter
loop iteration. At the \texttt{update\_context} intrinsic call, what
will constant-propagation know about \texttt{pc}?

One possible solution to this dilemma is to write the interpreter
control-flow with \emph{two} backedges, one for the taken- and one for
not-taken case. This way, the next PC is always constant at any given
static program point, and the interpreter's conditional branch becomes
the conditional branch in the compiled code. However, this approach
falls short: it is vulnerable to tail-merging optimizations when the
interpreter itself is compiled\footnote{We considered investigating
intrinsics or other optimization directives in the interpreter source
to prevent this optimization from breaking the \weval{} transform, but
in the end, we decided this was a philosophical dead-end: it is better
for the transform to work for \emph{any} code, optimized in \emph{any}
way (as far as practical). The calls to intrinsics will never be
optimized away when compiling the interpreter, because they are
external/imported functions; that is all that is necessary for
correctness.}, and it does not scale to opcodes with a dynamic number
of targets (from e.g. switch statements). In essence, we cannot reify
all control flow paths as branches in the interpreter if we do not
have a static number of paths for one opcode.

Instead, we introduce another intrinsic to allow \emph{splitting
context on values} (in the partial evaluation literature, this is
known as \emph{``The Trick''}~\cite{jones96}). The idea is that rather
than a scalar context (e.g., an interpreter PC), we add a sub-context
index, so specialization maps key on $\langle$ basic block, context,
value $\rangle$. We add an intrinsic:

\begin{lstlisting}[language=C++,numbers=none]
int32_t specialized_value(int32_t value, int32_t low, int32_t high);
\end{lstlisting}

\noindent that specifies a range of $N$ possible values, and passes
through a run-time value. At the intrinsic callsite, the block
specialization generates control flow to $N$ blocks, branching at
\emph{runtime} on \texttt{value}, then constant-propagating at
\emph{compile time} in each specialized path. The net result is that
as long as we have a statically-enumerable list of possible values for
a ``next PC,'' we can support arbitrary control flow operators in
bytecode such as \texttt{switch}.

\subsection{Maintaining Static Single Assignment (SSA) Form}

There is one optimization that is critical to grant the \weval{}
transform acceptable performance in practice. An SSA-based IR has the
key invariant that a value can be used only in the subtree of the
dominance tree below its definition -- that is, in a block that is
dominated by the block where it is defined. This invariant ensures
that the value is always defined before it is used during program
execution. Because the result of the \weval{} specialization transform
is a control-flow graph that resembles the \emph{interpreted
program's} control-flow graph rather than that of the interpreter, the
def-to-use relationships in the IR of the interpreter body may no
longer satisfy this invariant when transcribed over to the specialized
function body. Thus, we must somehow repair the SSA, or ensure by
construction that we do not violate this invariant.

A simple solution is to run the \weval{} transform on a restricted
form of SSA that uses values only in the blocks in which they are
defined, and otherwise passes all values across control-flow edges
explicitly with $\phi$-nodes (or equivalently, block parameters). This
guarantees correctness because it trivially removes any dependence on
inter-block dominance relations. However, while this solution is
correct, it leads to very high transform cost and overhead (in our
experiments, up to a 5x increase in block parameter count, yielding
very slow compilation of the result).

We implement an analysis that finds a ``minimal CFG cut'' across which
all values need to be made explicit block parameters. Intuitively,
this cut is around where paths from different contexts may merge,
e.g., the interpreter backedge: these points are where subgraphs of
the original CFG are ``glued together'' to form a new overall shape.

This analysis operates by, in a fixpoint loop, for each basic block,
finding the ``highest same-context ancestor'' (HSCA) in the dominance
tree. Each block flows its HSCA outward on CFG edges; blocks that have
update-context intrinsics have themselves as HSCA. If an HSCA flows
into a block and does not dominate that block, the block becomes a
cut-point and becomes its own HSCA. Otherwise, a block's HSCA is the
domtree-join (lowest common ancestor) of all inbound HSCA values. Once
we find all cut-points, we update these blocks to have block
parameters for all live values flowing in.

\subsection{Interface: Semantics-Preserving Specialization}
\label{sec:weval-semantics-preserving}

From the point of view of an interpreter and language runtime, how do
we integrate a transform that operates on the interpreter itself,
seemingly from outside the system?  Furthermore, how do we reason
about what the interface to this fragment of specialized code is, and
how we can integrate it, i.e. invoke it in place of the original
interpreter?

The key abstraction we provide is \emph{semantics-preserving
specialization}. The user of the \weval{} tool can make
\emph{specialization requests} that reference a function (e.g., a
generic interpreter) and include some constant arguments to that
function. The request causes the partial evaluator tool to generate a
new, specialized function. Each function argument is named in a
specialization request with one of three modes: \emph{Run-time},
\emph{SpecializedConst(value)}, or \emph{SpecializedMemory(data)}. The
first means the value is not known at compile time (no
constant-folding occurs), and the latter two specialize on either a
constant value or constant data at the given pointer, respectively. In
essence, the specialization request makes the \emph{promise} that the
function parameter or the memory contents will have those values at
invocation time: the semantics-preserving specialization is with
respect to this promise. In order to retain function-pointer type
compatibility, each specialized function continues to have parameters
even for specialized arguments. The specialized function body simply
ignores these parameters.

There are two general ways this API could be integrated into a system:
\emph{within} the execution universe of the program undergoing
specialization, or \emph{outside} of it. Both are reasonable for
different design points. An interpreter that already has a separable
frontend to parse and create bytecode might prefer to invoke \weval{}
``from the outside,'' appending new functions to an image of the
runtime. On the other hand, when adapting an existing interpreter with
no clear phase separation, it might make more sense to request a
specialization ``from the inside,'' directly providing data from the
heap and receiving a function pointer in return. This could operate at
run-time, with a JIT-compilation backend, or it could operate in a
\emph{snapshot} workflow: enqueue specialization requests, snapshot
the program with its heap, append new functions to the snapshot, and
restart. In our Wasm-based prototype, we take this latter approach,
building on top of the Wizer~\cite{wizer} snapshot tool. Note,
however, that this is not fundamental to the \weval{} transform.

When integrated into a Wasm-snapshot build workflow, the top-level
interface to our tool is a function that has a signature like the
following (slightly simplified):

\begin{lstlisting}[language=C++,numbers=none]
template<typename... Args>
request_t* specialize(func_t* result, func_t generic, Args... args);
\end{lstlisting}

This enqueues the ``request'' at a well-known location in the Wasm
heap so that the \weval{} tool can find it; when the Wasm module
snapshot is processed, the function pointer at \texttt{result} is
updated to point to the appended function.

The integration into an interpreter then requires one to: (i) enqueue
specialization requests when function bytecode is created; (ii) store
a specialized-code function pointer on function objects; and (iii)
check for and invoke this function pointer.\footnote{This is usually a
conditional, and the original interpreter may still be present if the
interpreted language allows, e.g., \texttt{eval()} at run-time, so not
every function may be specialized -- but this is outside the scope of
the \weval{} tool itself.} We will see objective measures of annotation
overhead, including this ``plumbing'' to weave the specializations
into the language runtime's execution, in the following sections.

\subsection{Generality Across IRs}
\label{sec:weval-ir-requirements}

We prototyped this transform on WebAssembly for pragmatic reasons (it
was the platform that spawned the need for our tool) but we believe
the transform is general. In brief, it will work on any IR and
platform given these requirements:

\begin{itemize}
  \item The possible control-flow edges need to be explicit -- for
    example, the IR cannot have a computed-goto feature with ``label
    address'' operators. Otherwise, it would not be possible to
    resolve block targets in specialization contexts ahead of time.
  \item The IR needs to support arbitrary, e.g., irreducible, control
    flow: when driven by specialized-on bytecode, i.e. user-controlled
    data, invariants of the original CFG such as reducibility may be
    lost. In our prototype on Wasm, where the output format can only
    represent reducible CFGs, we implement special lowering for
    irreducibility.
  \item The platform and tool interface together need to have a way to
    expose ``constant memory'' to the transform. In our prototype, the
    interface allows specifying a function argument as ``pointer to
    these constant bytes'' (e.g., bytecode or interpreter
    configuration data), and this works from inside the Wasm module,
    referring to bytes in the Wasm heap snapshot. However, one could
    also imagine an externally-driven interface where the data is
    provided separately.
\end{itemize}

\section{Handling Interpreter State Efficiently}
\label{sec:state}

\begin{wrapfigure}{r}{0.5\textwidth}
  \centering
  \includegraphics[width=0.5\textwidth]{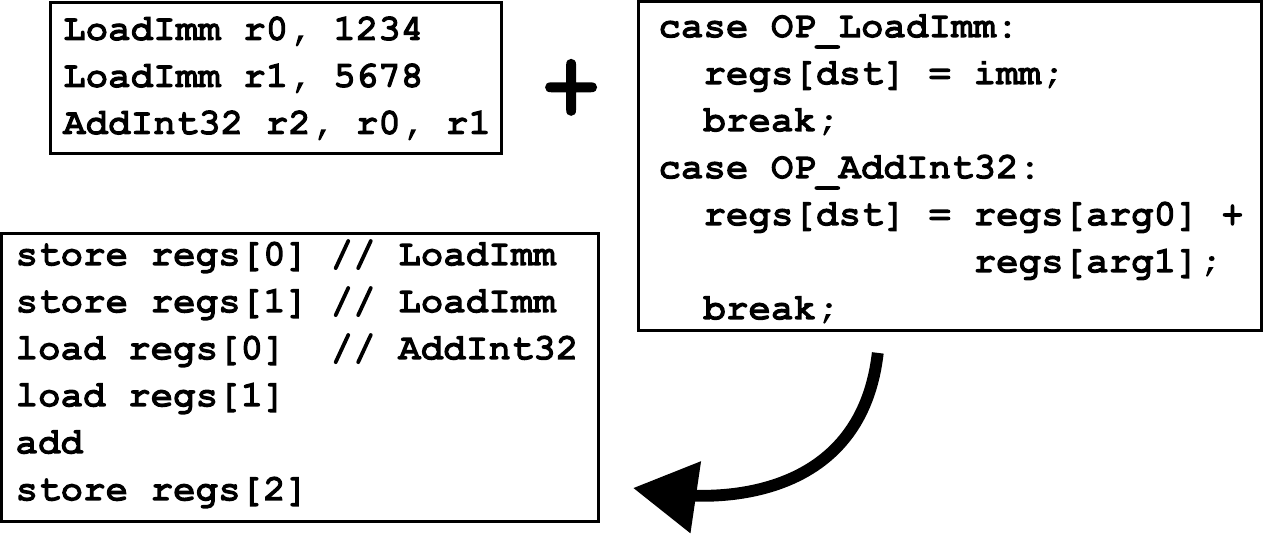}
  \caption{Partial evaluation by itself removes dispatch overhead, but
    preserves load/store semantics of interpreter state data
    structures, leading to inefficiency.}
  \label{fig:state-not-optimized}
\end{wrapfigure}
As it stands so far, our partial evaluator can eliminate an
interpreter's \emph{dispatch overhead} by pasting together the parts
of the interpreter's main loop that implement each opcode. However,
these opcode implementations will still likely contain dynamic
indirection to access the interpreted program's state. This is another
source of overhead that differentiates interpreted execution from
fully optimized compiled execution, and we wish to eliminate it as
well.

As a simple example, consider a bytecode for a virtual register-based
interpreter, together with opcode implementations, in
Fig.~\ref{fig:state-not-optimized}. If we were to take the Futamura
projection of this interpreter over the bytecode, we might obtain a
compiled result like that in the figure.

The \texttt{regs} array accesses compile to loads and stores to
offsets in the interpreter's state. A good alias analysis, combined
with redundant load elimination, dead store elimination, and
store-to-load forwarding optimizations, \emph{might} be able to
disambiguate these loads and stores. However, a realistic interpreter
might have other features that interfere with this: for example, calls
to other functions. These functions may not access \texttt{regs}, but
this cannot be proved intraprocedurally. Ideally we would like to
indicate some other way to the partial evaluator that these values can
be stored in true locals (i.e., SSA values in the \weval{} transform
result) rather than memory.

\subsection{Virtualized Registers}
\label{sec:state-regs}

In our tool, we allow the interpreter to communicate this intent via
intrinsics. Specifically, we provide the intrinsics
\texttt{load\_register(index)} and \texttt{store\_register(index,
  value)} that are semantically equivalent to loads and stores to a
hidden array within the specialized function. The \texttt{index}
parameter must always be a constant (perhaps loaded from the constant
bytecode) during specialization. (See \S\ref{sec:casestudy-toy} for an
example that uses these intrinsics.) The specialization transform
carries a map of indices to SSA values, and translates these
intrinsics appropriately, reconstructing SSA (by inserting block
parameters at merge points) where needed.

\subsection{In-Memory State: Locals and Operand Stack}

Non-escaping locals provide an important primitive, but interpreters
sometimes have state that is necessarily exposed to the rest of the
runtime. For example, in a garbage-collected platform, a GC might need
to inspect an array of local-variable values in order to mark them as
rooted or update them after a compaction.

As above, we wish to lift the original in-memory storage to SSA when
possible. However, these values will need to be written back to memory
at certain points, and their new values reloaded. To support this, we
provide two state abstractions that build on virtualized registers,
but carry both the value \emph{and} a canonical in-memory
address. This kind of state operates like a write-back cache: the
transform will perform true loads when necessary, and will generate
stores at ``flush'' intrinsics. The intrinsics for in-memory locals
are \texttt{read(index, address)}, \texttt{write(index, address,
  value)}, and \texttt{flush()}.

Beyond indexed locals, many interpreters also present a \emph{stack
VM} abstraction with opcodes that push and pop operands and
results. To support this, we also provide \texttt{push(address,
  value)}, \texttt{pop(address)}, \texttt{read\_stack(depth,
  address)}, and \texttt{write\_stack(depth, address, value)}
intrinsics. These perform an abstract interpretation of stack state,
falling back to true loads when needed, and generating stores at
flush-points as above.

Note that some care must be taken to ensure that \texttt{flush()} is
invoked wherever the in-memory state might be observed. In our
SpiderMonkey adaptation (\S\ref{sec:casestudy-spidermonkey}) we built
a C++ RAII mechanism to ensure this (exposing the ability to call the
rest of the runtime only after interpreter state is flushed). Any
interpreter that opts into these intrinsics will need to take care
that when in-memory state is observable, a flush has occurred. Other
design points might also be possible: for example, a new intrinsic or
mode in our tool that flushes at every callsite, or that tracks
escaped pointers to the state in some other way.

\subsection{Discussion: Semantics-Preservation and Polyfills}

These intrinsics differ from the initial function-specialization
transform in \S\ref{sec:weval} in two ways: (i) they grant the
\weval{} tool permission to diverge in semantics in controlled ways
(\emph{lazy flushing} of in-memory state with user-denoted sync
points), and (ii) they are not simply intrinsics that can be removed
(``hints'') but must be replaced/polyfilled for ordinary execution of
the original function body to work. This permission to diverge is
fundamental for performance: the memory operations become a severe
bottleneck otherwise.

\begin{wrapfigure}[15]{r}[-2pt]{0.5\textwidth}
  \vspace{-\baselineskip}
  \centering
  \includegraphics[width=0.5\textwidth]{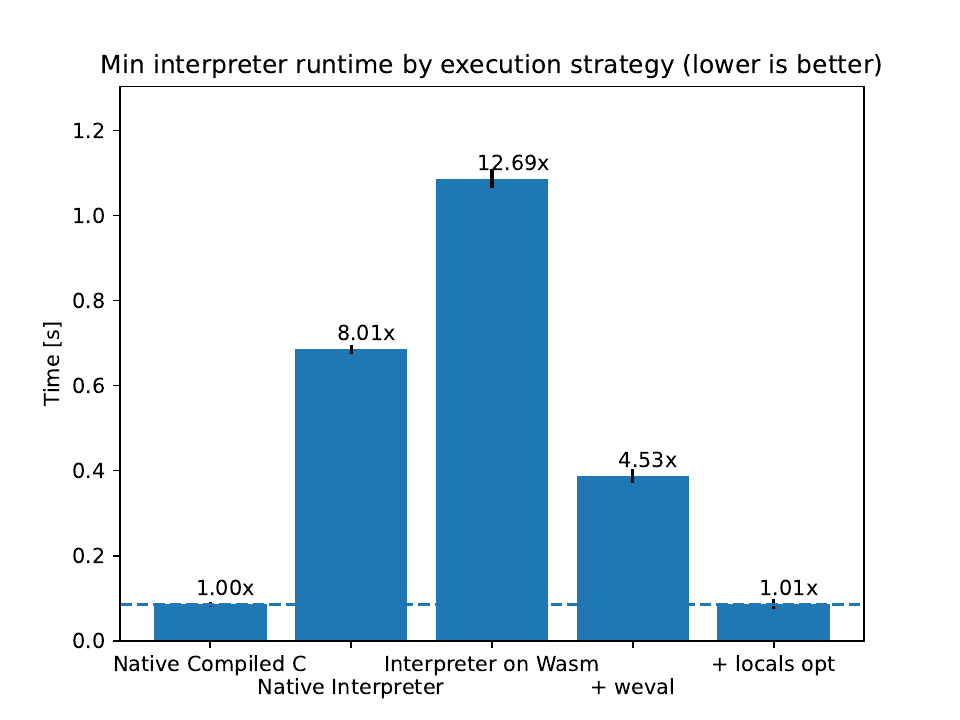}
  \caption{Benchmark of the loop program with different execution strategies.}
  \label{fig:min-benchmark}
\end{wrapfigure}

Note that we have carefully designed the signatures so that polyfills
are possible: the in-memory state intrinsics take canonical address
arguments and can thus fall back to true loads and stores. The
virtualized register intrinsics (\S\ref{sec:state-regs}) could be
rewritten to loads and stores to an array allocated on the stack. For
pragmatic reasons we have no implemented these polyfills, and instead
we generate two separate versions of the interpreter body function
with and without state intrinsics, but this is not fundamental.

\section{Case Study: Minimal Toy Interpreter}
\label{sec:casestudy-toy}

To give a feel for \weval{}, we integrate it first into a minimal example
interpreter---a small 64-bit unsigned integer register machine named
\textit{Min}. Min has 10 instructions that operate on a program
counter $pc$, an array of indexed registers $registers$, and a special
accumulator register $acc$.  Except for the \verb|JMPNZ| instruction,
the machine reads the instruction, increments the $pc$, executes the
instruction, and returns to the top of the interpreter loop.  The
interpreter loop is summarized
in~\Cref{fig:min-interpreter}.

\begin{figure}
  \captionsetup{width=0.45\textwidth}
  \centering
  \begin{minipage}[t]{.5\textwidth}
\begin{lstlisting}[language=C]
uint64_t Execute(uint64_t *program) {
  uint64_t accumulator = 0;
  uint64_t registers[256] = {0};
  uint32_t pc = 0;
  PUSH_CONTEXT(pc);  // NEW
  while (true) {
    uint64_t op = program[pc++];
    switch (op) {
    case LOAD_IMMEDIATE: {
      accumulator = program[pc++];
      break;
    }
    case STORE_REG: {
      uint64_t idx = program[pc++];
      registers[idx] = accumulator;
      break;
    }
    // ...
    case ADD: {
      uint64_t idx1 = program[pc++];
      uint64_t idx2 = program[pc++];
      accumulator = registers[idx1] +
        registers[idx2];
      break;
    }
    // ...
    } // end switch
    UPDATE_CONTEXT(pc); // NEW
  } // end while
  POP_CONTEXT(); // NEW
}
\end{lstlisting}
\caption{\emph{Min} bytecode interpreter in C. The listing is
  shortened for this paper but in total is 63 lines of code. Lines
  marked \texttt{NEW} are the added \weval{} annotations.}
\label{fig:min-interpreter}
\end{minipage}%
  \begin{minipage}[t]{.5\textwidth}
    \begin{lstlisting}[language=C++]
#define REG_AT(idx) (IsSpecialized ? \
      read_reg(idx) :                \
      registers[idx])
#define REG_AT_PUT(idx, val)   \
  if (IsSpecialized) {         \
    write_reg(idx, val);       \
  } else {                     \
    registers[idx] = val;      \
  }

template <bool IsSpecialized>
uint64_t Execute(uint64_t *program) {
  // ...
      REG_AT_PUT(idx, accumulator);
  // ...
      accumulator = REG_AT(idx);
  // ...
}
  \end{lstlisting}
  \caption{We modify the macros to read and write registers to
    conditionally use \wevals{} register intrinsics. For non-fundamental
    reasons, we currently don't polyfill the intrinsics in our tool in
    non-specialized versions of the function, so we need to generate
    two versions of the interpreter function: one using the
    intrinsics, and one with a conventional register array. In order
    to create both alternatives, we use C++ template specialization to
    ensure this choice is made when the interpreter is compiled.  In a
    pure C-based interpreter, one could put the interpret function in
    a separate file, redefine the macros twice (once for intrinsics
    and once without), and include the function body in both cases.}
\label{fig:registers-wevaled-min-interpreter}
  \end{minipage}
\end{figure}

The first step to \wevaling{} an interpreter is adding a
\textit{context} annotation. To specialize a bytecode interpreter, we
use the program counter---the \verb|pc|---as the context. As the
annotation only has meaning when the program is being partially
evaluated, we invoke the annotations with macros that are
conditionally defined only in a build for \weval{}.

Just these annotations alone will confer a performance boost; they
unroll the interpreter loop into guest language control flow and allow
for \wevals{} optimizer ``see through'' the interpreter into the guest
language. We can, however, do better: we can also use \wevals{}
interpreter state optimizations (\S\ref{sec:state}).

To allow \weval{} to optimize $registers$ and remove loads/stores, in
Fig.~\ref{fig:registers-wevaled-min-interpreter} we replace direct
array accesses like \texttt{registers[idx]} with macros
\texttt{REG\_AT(idx)} and \texttt{REG\_AT\_PUT(idx, value)} and define
them to use \texttt{read\_reg(idx)} and \texttt{write\_reg(idx,
  value)} in a variant of \texttt{Execute} passed to the partial
evaluator. Now we can unroll the ``infinite register'' bytecode that
reads from and writes to an array into direct SSA dataflow. This
avoids touching memory \textit{and} gives more information to the
optimizer.

As a benchmark, we write one program in Min that computes the sum of
all integers from 0 to 100 million and prints it to
\textit{stdout}. In Fig.~\ref{fig:min-benchmark} we show performance
of a Min program running on the C++ interpreter running on the host
platform directly (as an x86\_64 program), on the interpreter compiled
to Wasm, and on that interpreter processed by \weval{} and with
interpreter-state optimizations applied (all Wasm variants running on
Wasmtime), all compared to the equivalent program written in
C.\footnote{We find that a sufficiently advanced optimizer, such as
the one present in Clang/LLVM, can completely unroll the loop into the
closed-form $n(n+1)/2$. Adding such an optimizer (for example,
Binaryen~\cite{binaryen}) is future work and not relevant to the main
claims of this paper. To keep the loop and local variables around for
the benchmark, we annotate with \texttt{volatile}.} We show that
\wevaling{} the interpreter beats native interpreter performance and
unrolling local variables yields still more speedup, coming within 1\%
of the performance of the equivalent program written in C and compiled
to native code.

Min is an excellent introduction to \weval{} because a first-year graduate
student learned about the idea of \weval{}, designed an interpreter, and then
directly applied \weval{} with minimal tutelage---all in the space of four hours.
This indicates to the authors that \weval{} is not only useful, but immediately
applicable to many extant projects.

\section{Case Study: SpiderMonkey JavaScript Interpreter}
\label{sec:casestudy-spidermonkey}

In this section, we present our most significant real-world use-case:
an application of our partial evaluator to the
SpiderMonkey~\cite{spidermonkey} JavaScript engine's interpreter, in
order to derive compiled code directly from the interpreter
semantics. This application of our tool has been merged into the
StarlingMonkey JS runtime~\cite{starlingmonkey} which embeds
SpiderMonkey to target Wasm-first platforms, where run-time code
generation (JIT) is not supported. \weval{}-based snapshot processing
is used to provide its ``ahead-of-time compilation'' (AOT) feature.

The SpiderMonkey JavaScript engine, running on a native platform
(e.g. x86 or ARM), has several interpreter tiers and several
JIT-compilation tiers. It does not have support for ahead-of-time
compilation. It has been ported to run within a WebAssembly
module~\cite{spidermonkey-wasi}; in this mode, it runs only with its
interpreter tiers, because Wasm does not support run-time
code-generation. SpiderMonkey supports inline caches (ICs) in a fully
interpreted mode using the \emph{Portable Baseline Interpreter}
(PBL)~\cite{pbl-post-2023}; we take this as our baseline. We augment
StarlingMonkey's compilation flow that uses the Wizer~\cite{wizer}
snapshotting tool: once a snapshot of the engine with the user
program's bytecode is taken, with function specialization requests
enqueued in the heap image, \weval{} processes those specialization
requests and appends new Wasm functions to the module.

There is one further hurdle we must address: IC bytecode is generated
only at run-time, when behaviors are observed, while our tool requires
bytecode to be present in the snapshot for AOT processing. To address
this dependency, we augment SpiderMonkey with an \emph{IC corpus}
mechanism that builds a pre-collected set of IC bodies into the binary,
available in a lookup table.

Stated succinctly, the key insight is: ahead-of-time compilation of
JavaScript is possible at this level because inline caches (ICs) allow
dynamism in semantics to be pushed to late-binding run-time data
changes (function pointer updates) rather than code changes.

\subsection{Changes to the Interpreter}

In order to permit the Portable Baseline Interpreter (PBL)'s two
interpreter loops -- for JS and IC bytecode -- to be partially
evaluated, several minor changes were necessary. First, we had to
ensure that one native function call frame (in the interpreter's
implementation language, C++) corresponded to one JS function or IC
stub call; this is what allows per-function specialization to
work. The interpreter was originally written to perform JS calls and
returns ``inline'', by pushing and popping JS stack frames as data
without making C++-level calls. We modified the interpreter to recurse
instead.

Second, as in the previous section, we added annotations to update
context, and to optimize the storage of interpreter state. We used
\wevals{} ``registers'' for CacheIR, which is a register-based IR; and
``locals'' and the virtualized operand stack for JS bytecode.

To handle several forms of non-local control flow, our modified
interpreter loop tail-calls (``restarts'') to a non-specialized
version of itself -- just for the active function frame -- in several
control flow situations that are nontrivial or inefficient to handle:
\texttt{async} function restarts, which would imply multiple function
entry points\footnote{It should be possible to either include a
\texttt{switch} at the beginning of async functions to handle this, or
more ambitiously, define new intrinsics that allow compiling
coroutine-like code to WebAssembly's stack-switching proposal; we have
not yet implemented either approach.}, and error cases (including
exception throws), to minimize compiled code size. The engine supports
all edge cases and retains 100\% compatibility (continues to pass all
tests), and only a negligible fraction of execution time is spent in
interpreted (non-specialized) code in our benchmarks.

Our patch to add these annotations and intrinsics amounted to
\texttt{+1045 -2 lines}, including a vendored \texttt{\weval{}.h}. The
changes to the interpreter function itself amount to 133 lines of
alternate macro definitions to swap in the intrinsic usages.

\subsection{Performance Results}
\label{subsec:spidermonkey-wasm-perf}

\begin{figure}
  \centering
  \includegraphics[width=1.0\textwidth]{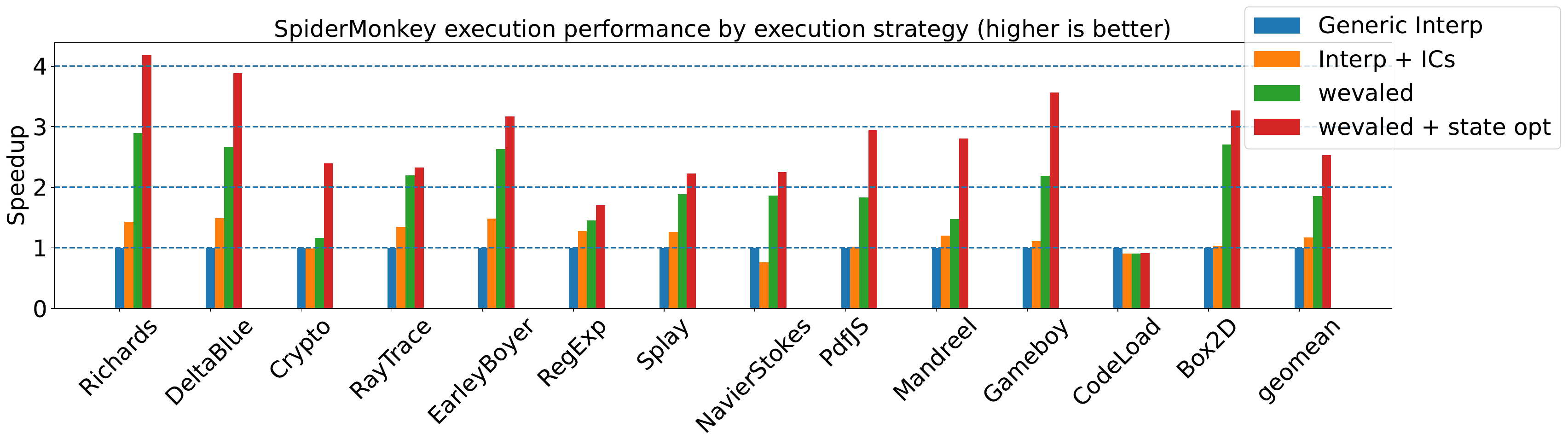}
  \caption{Performance results of Octane benchmark suite on SpiderMonkey
    engine, with interpreter in a Wasm module (without and with ICs),
    and \weval{}-compiled code (without and with interpreter-state
    optimizations).}
\label{fig:spidermonkey-main}
\end{figure}

In Fig.~\ref{fig:spidermonkey-main}, we show the performance of our
modified SpiderMonkey engine on the Octane benchmark
suite~\cite{octane}, reporting throughput (inverse run time, i.e.,
speed) data for these configurations, all as Wasm modules running on
Wasmtime~\cite{wasmtime}:

\begin{itemize}
  \item \emph{Generic Interp}: default (generic) interpreter;
  \item \emph{Interp + ICs}: Portable Baseline (PBL) interpreter
    with inline caches;
  \item \emph{\wevaled{}}: AOT-compiled JS via partially-evaluated PBL
    interpreter;
  \item \emph{\wevaled{} + state opt}: same, with state optimizations
    (\S\ref{sec:state}) -- our final configuration.
\end{itemize}

The speedup of our approach is properly seen as the delta from
\emph{Interp + ICs} to \emph{\wevaled{} + state opt}. This ratio is a
$2.17\times$ speedup; up to $2.93\times$ on the best benchmark
(Richards) and above $2\times$ in all cases except RegExp (which
depends heavily on the regular expression engine's interpreter loop
which we have not modified) and CodeLoad (which tests the engine's
code loading rather than execution speed).

Our use of interpreter state optimizations was motivated by the
observation that loads and stores to locals and the operand stack are
quite hot. By making use of these intrinsics, from the \emph{\wevaled{}}
configuration to \emph{\wevaled{} + state opt}, we see a $1.37\times$
speedup. Without this optimization, a number of benchmarks see almost
no speedup at all (Crypto, Mandreel). Across all of Octane, the
virtualized stack intrinsics elide 84\% of 639K loads and 76\% of 563K
stores; the local intrinsics elide 14\% of 149K loads and 5\% of 74K
stores. (Pushes and pops happen at every opcode, while JS locals are
accessed less frequently and there are more often GC safepoints in
between that force values back into memory.)

\subsection{Comparison to Native Execution}

\begin{figure}
  \centering
  \includegraphics[width=1.0\textwidth]{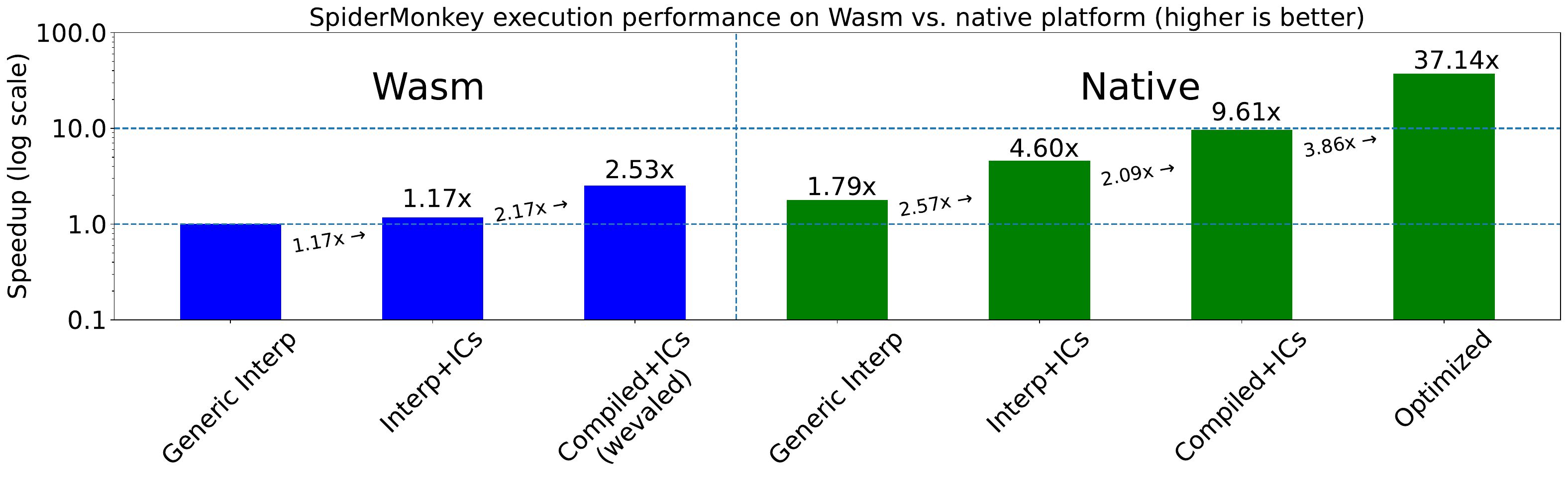}
  \caption{SpiderMonkey configurations running on top of a Wasm engine
    vs. SpiderMonkey as a native build on the same system. This shows
    how (i) inline-cache fastpaths, (ii) compilation of JS bytecode
    and inline caches separately, and (iii) optimized compilation of
    both together (native only) result in progressive speedups.}
\label{fig:spidermonkey-native}
\end{figure}

In order to judge the relative speedups attained by \weval{}, and also
eventual upper bounds, it is interesting to compare analogous execution
strategies in a native-code configuration, with JIT backends. Note that
we do not intend to compare \weval{}'d code inside a Wasm module
\emph{directly} to the native code -- it encounters some overhead due to
the Wasm sandbox -- but rather, the progressive ratios of each step on
the two platforms.

Fig.~\ref{fig:spidermonkey-native} shows three of the four
configurations from \S\ref{subsec:spidermonkey-wasm-perf} on Wasm
alongside four configurations running natively (e.g. directly on
x86\_64) on the same system:

\begin{itemize}
\item \emph{Generic Interp}: the same generic interpreter as in the
    Wasm case (\texttt{js --no-ion --no-baseline --no-blinterp});
\item \emph{Interp + ICs}: SpiderMonkey's baseline interpreter, which
    interprets JS bytecode but runs \emph{compiled} IC stubs (\texttt{js
    --no-ion --no-baseline});
\item \emph{Compiled + ICs}: SpiderMonkey's baseline compiler, which
  compiles JS bytecode and ICs, comparable to the optimization level
    of \wevaled{} code (\texttt{js --no-ion});
\item \emph{Optimized}: a fully optimized combination of JS bytecode
    with type-specialized cases inlined, yielding maximal performance
    (\texttt{js} default native engine).
\end{itemize}

We label speedup ratios between each successive pair of configurations.
A few interesting comparisons can be made. First, by observing the
second to third bar on each side, this plot shows that \weval{} attains
a similar speedup over the next lower tier (interpreter with ICs) as the
native baseline compiler does. In both cases, we are removing the
interpreter overhead but retaining runtime binding of behavior via IC
stubs.

Second, we see that fully optimized native JIT execution is a
significant speedup ($3.86\times$) over baseline compilation. Note that
the \emph{Optimized} configuration ``pulls out all the stops'' and, in
particular, takes advantage of being a \emph{JIT compiler}: it
type-specializes code. As we argue in \S\ref{sec:future}, we believe
there is a path for our AOT-based approach to adopt profile-guided
inlining in a safe, principled way, possibly closing this gap.
Nevertheless, the gap remains today.

Third, however, the overall speedup of the ``baseline compiler-like''
configurations -- first bar to third bar -- is still somewhat behind in
\weval{}: $2.53\times$ over the generic interpreter, vs.  $5.37\times$
on native.  In principle there is no difference between the
optimizations that both configurations are capable of, and in fact
profiling and examination of generated code largely bears this out: both
are ``baseline compilation'' producing a skeletal compilation of JS
bytecode that invokes ICs and plumbs values between them, and
straightline compilations of IC opcodes.  We believe the remaining
inefficiencies lie mostly in the IC invocation efficiencies: the native
baseline compiler can tightly control ABI and register allocation,
keeping hot values in pinned registers and effectively doing
interprocedural register allocation between the JS function body and
ICs. In contrast, on a Wasm platform, control-flow integrity (CFI)
checks make indirect calls much slower.

\subsection{Code Size}

The Wasm module containing the entire SpiderMonkey JavaScript engine
contains 8~MiB of Wasm bytecode initially, in 18080 functions. After
AOT compilation with \weval{} of the entire Octane benchmark suite
(7.5MiB or 337KLoC of JS) together with the pre-collected corpus of
2320 ICs, there is 52~MiB of Wasm bytecode, with 5212 new functions
from JS function bodies and 2320 new IC-stub functions. With more
optimization work in our tooling, including our Wasm compiler backend,
we believe the size of generated bytecode could be decreased
substantially.

\subsection{Transform Speed}

Compiling the above Wasm module takes, in total, 350 seconds of CPU
time (44.16 seconds wallclock parallelized over specialization
requests on a 12-core machine). This indicates a compilation speed of
slightly under 1KLoC/second of JavaScript source. We believe this
could be improved with further work: our Wasm compilation backend has
not been heavily optimized. In order to improve compilation times in
practice, we have added a cache that keys on input Wasm module hash
plus the function specialization request's argument data; in practice,
this works well to avoid redundant work for the unchanging AOT IC
corpus, and helps with incremental compilation during development as
well.

\section{Case Study: PUC-Rio Lua Interpreter}
\label{sec:casestudy-lua}

To demonstrate generality of the tool across multiple real-world
bytecode interpreters, we ported the original (PUC-Rio) Lua
interpreter, with which no author had any familiarity, to Wasm and
applied \weval{}-based partial evaluation in the space of three hours. We
split the process into the following chunks:

\textit{Support Wasm.} Porting the interpreter to Wasm took
approximately 45 minutes. For simplicity, we stubbed out (i.e. removed
the source and added calls to \verb|abort()|) some Linux-specific OS
library functions; we also stubbed out exception handling because it
uses the \texttt{setjmp} and \verb|longjmp| C
functions\footnote{Unfortunately, many WebAssembly runtimes do not
\textit{yet} support \texttt{setjmp} and \texttt{longjmp} but support
is expected to land soon with the exception-handling
extension~\cite{wasm-exceptions}. For now, projects such as Emscripten
handle exceptions by calling into the host JavaScript runtime and
leaning on JavaScript exceptions.}.

\textit{Support Wizer-based snapshotting.} After the initial Wasm
port, we spent another 45 minutes adding snapshotting. This involves
adding approximately 30 lines of C code near the C \verb|main|
function to expose two functions: \verb|wizer_init| and
\verb|wizer_resume|. The initialization function runs the top-level
Lua module and pushes that module's \verb|main| function (a convention we
arbitrarily chose) onto the call stack. The resume function---the new C
\verb|_start|---calls this function.

\textit{Specialize functions.} Supporting function specialization requires
adding two pointer-sized fields to Lua's function object (\verb|Proto|) struct:
a specialized function pointer $spec$ and a \weval{} request pointer $req$.
We create and fill in $req$ when the function object is created in the parser.
It must exist somewhere in the heap so that \weval{} can find it, and we retain it
so that it can be freed later on function destruction. (It is possible to
instead use a side-table, but we chose to keep the implementation simple.)
When we make the \weval{} request, we also pass it the address of $spec$ field for
\weval{} to fill in later, after Wizer has run.

We tested this step before we added annotations to the interpreter: at
this point, \weval{} specialization should produce the same interpreter
function as output, because no context-specialization occurs. We also
modified the interpreter function (\verb|luaV_execute|) signature to
pass in a bytecode parameter for specialization.

\textit{Annotate interpreter.} PUC-Rio Lua uses macros instead of manual code
duplication to implement much of its interpreter control-flow. This makes
modifying the interpreter straightforward: we add a \verb|push_context| to the
top of the interpreter and an \verb|update_context| to back edges.

\textit{Change call path.} In order to reap the benefits of our
specialized function pointers, we must call them. Lua has only two
ways to call a managed function (outside the interpreter and inside
the interpreter) and we modified both to call $spec$ if it had been
filled in. We also ensured that the interpreter calls itself for each
Lua call, rather than handling the call opcode ``inline''.

After a short period of debugging, we had a working ahead-of-time
compiler for Lua. Some trivial interpreter-heavy benchmarks produce
the expected results, showing a $1.84\times$ speedup. The resulting
source tree has a diff in Lua C/header files (excluding \wevals{} and
Wizer's headers, and build-system tweaks) of \texttt{+173 -57}
lines. This includes the initial port to Wasm. Future work includes
calling intrinsics to lift local variables or stack variables to Wasm
locals. We expect this to take not significantly more time for
programmers who are already familiar with the code.

\section{Related Work}

\noindent\textbf{Partial Evaluation:} There is a rich pre-existing
literature on partial evaluation, going back at least to
Futamura~\cite{futamura71,futamura99}. Jones~\cite{jones96} provides a
comprehensive overview of the field. Several aspects of the \weval{}
transform, such as its use of value specialization (``The Trick''),
are standard techniques for partial evaluators; and others implement
optimizations such as virtualized handling of interpreter state (e.g.,
PyPy~\cite{pypy}) as well. Our particular tool differs primarily in
targeting WebAssembly, and possibly in its particular basic block- and
SSA-oriented specialization transform.

\noindent\textbf{DyC}~\cite{dyc-1999,dyc-2000} is a run-time
optimization system for C that performs partial evaluation. It targets
use-cases where run-time data could be used to greatly simplify or
specialize a program's logic: for example, simulators (for a
particular configuration) or numeric code (for known dimensions or
parameters), in addition to interpreters. It provides annotations
(e.g., \texttt{make\_static()} to perform ``The Trick'' of value
specialization) and performs binding-time analysis; the work
demonstrates the ability to unroll interpreter loops, as we
do. However, the system appears to be significantly more complex,
relying on heuristics and analysis to handle interprocedural
specialization, overlapping specialization regions, caching of
specializations, and more. In contrast, our tool (\weval{}) is only 5K
LoC, and provides a comprehensible semantic model for its
transform.

\noindent\textbf{GraalVM and Truffle}~\cite{graal-pldi2017} are a JIT
compiler backend and language runtime framework (respectively) that
perform the first Futamura transform. The Truffle ecosystem supports
high-performance execution of
Ruby~\cite{cseaton-thesis,graal-pldi2017},
JavaScript~\cite{graal-pldi2017}, and other popular languages.

Truffle users rely on a Truffle-provided framework that supports
AST-walking interpreters. Users' AST node classes must inherit from
Truffle classes. In other words, the interpreter must be developed
specifically for Truffle. The engine also includes support for run-time
optimization and de-optimization based on changing or violated
assumptions, which is very useful for dynamic language support. The
tradeoffs come in terms of significant added complexity, and long
warmup times. More recently, Truffle has been used to build
interpreters for bytecode (e.g., TruffleWasm~\cite{trufflewasm}),
though by building AST nodes for bytecode instructions.

GraalVM also supports a ``Native Image'' feature that shares some
motivation (startup latency, etc) with the Wizer~\cite{wizer}
WebAssembly snapshotting tool (used as part of the workflow in our
case studies), but it goes further: it runs static initializers, then
uses a closed-world assumption and a loop of (optionally
context-sensitive) points-to analysis and optimization to specialize
Graal IR.  Then it produces a small native binary.
Taken together with a Truffle-based interpreter, this presents another
way to build an AOT compiler for a language from an interpreter.

\noindent\textbf{PyPy}~\cite{pypy} is a Python implementation built on
top of the RPython meta-tracing JIT compiler infrastructure. RPython
has a similar API to our annotation infrastructure. Constructing an
instance of the \verb|JitDriver| class requires annotating which
variables are constant for the execution of a particular instruction
and which are not; \verb|JitDriver.jit_merge_point| unrolls the
interpreter loop using an arbitrary context (ordinarily the \verb|pc|
value). RPython also allows for optimizing interpreter state, as we
do. RPython is different, however, because (i) interpreters must be
written in RPython, a restricted subset of Python 2.7, and (ii) the
interpretation and compilation strategy is based on a linear program
trace instead of an entire method or CFG.  Additionally, it is not
possible to use RPython for AOT compilation.

\noindent\textbf{BuildIt}~\cite{buildit} is a C++ library for partial
evaluation of C++ programs.  It provides similar annotations in the
form of templated types---\verb|static_var| and \verb|dyn_var|---for
partitioning variables into compile-time constants and run-time data,
respectively. BuildIt generates C++ code. With BuildIt, it is possible
to do similar unrolling of interpreter loops and promotion of local
variables, but it is easier to go wrong; mis-use of \verb|static_var|
(analogous to our \verb|push_context|) can lead to semantic
differences other than performance. Additionally, unlike \weval{},
BuildIt relies on the user to provide a C++ compiler to compile the
generated code.

\noindent\textbf{Lightweight Modular Staging (LMS)}~\cite{lms,
  lms-essence} is a library developed by Rompf and Odersky for partial
evaluation of Scala programs. It has been used to great effect to,
among other things, compile SQL queries to efficient
code~\cite{legobase}. Like BuildIt, it is general-purpose and requires
annotations like \verb|Rep| (comparable to BuildIt's
\verb|static_var|). Unlike BuildIt, it can target more language
backends. Using this library requires using Scala as the host
language.

\noindent\textbf{Deegen}~\cite{deegen} aims to generate a fast
interpreter, baseline JIT, and---eventually---an optimizing JIT from a
description of language semantics.  Deegen provides a C++ DSL for
describing interpreter semantics in the form of an infinite register
bytecode VM. It has APIs for defining opcodes, type-specialized
variants of opcodes, inline caches, a type lattice, ``slow paths'',
and more. Similarly to PyPy, it cannot be used for AOT compilation; it
intentionally burns constants into the generated code for better
performance.

\noindent\textbf{SemPy~\cite{sempy} and Static Basic Block
  Versioning~\cite{sbbv} (SBBV)} are works by Melançon, Feeley, and
Serrano that seem to be building toward a similar goal of deriving a
compiler from interpretable semantics using context-sensitive dataflow
analyses and partial evaluation. SemPy records language semantics in
canonical, interpreter-like form, but this form is developed
explicitly for the purpose, i.e., is not a pre-existing
interpreter. SBBV is a complementary technique to a semantics with
many dynamic type checks: it is an algorithm for finding a set of
type-specialized contexts ahead-of-time to generate code for (without
knowing what the real types will be), while applying a technique to
limit combinatorial blowup. This is an alternative to our approach to
dynamic types in SpiderMonkey: we observe that AOT compilation is possible when
inline caches (ICs) are used if all type-driven dynamism is pushed to
late-binding (run-time-filled) IC callsites, while SBBV allows
optimization beyond that level, permitting type-specialized variants
of code with strongly typed values fully unboxed. These are
complementary, and an SBBV-like algorithm could in theory be applied
to a \wevaled{} interpreter body with intrinsics denoting
type-specialization points.

\noindent\textbf{LuaAOT}: We apply \weval{} to a Lua interpreter to
AOT-compile Lua bytecode. LuaAOT~\cite{lua-aot} is a purpose-built AOT
compiler framework that, superficially, works similarly: it compiles
bytecode by pasting together portions of the interpreter
loop. However, its algorithm operates at the source (interpreter in C) level.
The work claims 20--60\% speedups ($1.25\times$--$2.5\times$) from a 500-LoC
implementation; in contrast, our Lua modification achieves
$1.84\times$ speedup (on top of WebAssembly, though in principle
\weval{} is not limited to Wasm, as we noted in \S\ref{sec:weval})
with a \texttt{+173 -57}-line diff.

\noindent\textbf{AOT JS Compilers:} While AOT compilation of
JavaScript is not the focus of our work, our evaluation on
SpiderMonkey does yield such a compiler. Several other options exist.
Hopc~\cite{hopc, hopc-performance}, Porffor~\cite{porffor}, Static
Hermes~\cite{static-hermes}, and Static
TypeScript~\cite{static-typescript} are all compilers that accept
JavaScript, TypeScript, or some annotated subset thereof, and produce
either native code or WebAssembly. The key distinction from our work
is that these compilers are explicit: they are written as code
transforms, not executable interpreters, and hence are harder to
validate, debug or extend than an interpreter-based solution. On the
other hand, they can perform optimizations in a more straightforward
way, potentially yielding higher ultimate performance.

\section{Future Work}
\label{sec:future}

\subsection{Profile-Guided Inlining and Semantics-Preserving Optimizations}

The level of optimization that the \weval{} transform is able to
provide in its current form (as a processing step on a program
snapshot) is limited by its ahead-of-time-only design goal: it cannot
optimize based on types in dynamic languages. To carry the goal of
automatically deriving a compiler further, one ought to be able to
derive type-specialization optimizations beyond ICs.

We believe that \emph{profile-guided inlining} is a principled way to
do this. One would start with an AOT compilation, observe and gather
statistics on IC callsite targets, and eventually recompile
indirect-call instructions into \emph{guarded inlined functions} (if
function pointer is X, inline body, else call-indirect). In this way,
just as for the basic \weval{} transform, \emph{semantics are fully
preserved}. The Winliner tool~\cite{winliner} prototypes this
optimization strategy. This parallels how SpiderMonkey's
WarpMonkey~\cite{warpmonkey} backend generates its optimizing compiler
input from inlined ICs.

Such inlining would remove the IC indirect-call overhead, and it
indirectly creates opportunity: it places \emph{boxing} and
\emph{unboxing} operations together in the same function body. Our
tool could then incorporate special optimizations that -- still
preserving semantics -- hoist type-checks upward, and eliminate
boxing-unboxing pairs. All of these operations can be written as
generic compiler rewrite rules -- for example, SpiderMonkey's boxing
is a form of NaN-boxing and so a partial-known-bits optimizer that
understands known tag bits and conditional checks on them should
achieve this~\cite{ebpf-knownbits,pypy-knownbits}. Overall, this is a
further step toward a goal of \emph{safe dynamic-language compilers},
where ``safe'' means that the only ``load-bearing'' semantics are
described in the interpreter body.

\subsection{Specializing at Run-time}

While our initial prototype is used to derive AOT compiler backends, we are
also interested in generating JIT backends. Nothing prevents the \weval{}
transform from operating at run-time.
Operating at run-time places stresses on the performance of the
specialization algorithm itself. In principle, the \emph{second}
Futamura transform could generate an efficient compiler directly from
the interpreter, without the specializer (the \weval{} tool) remaining
at all. This is out of reach today, but remains an interesting
end-goal.

\bibliography{paper}

\end{document}